  \documentclass[12pt,preprint]{aastex}

  \usepackage{epsfig}

\newcommand{\ea}{et al.}

\newcommand{\kms}{\>{\rm km}\,{\rm s}^{-1}}
\newcommand{\pc}{\>{\rm pc}}

\newcommand{\mpc}{\>{\rm Mpc}}
\newcommand{\m}{\>{\rm m}}

\newcommand{\ergs}{\>{\rm ergs}\,{\rm s}^{-1}}
\newcommand{\cm}{\>{\rm cm}^{-2}}

\newcommand{\mum}{\>{\mu {\rm m}}}
\newcommand{\kev}{\>{\rm keV}}
\newcommand{\jy}{\>{\rm Jy}}
\newcommand{\mjy}{\>{\rm mJy}}

\newcommand{\lsun}{\>{\rm L_{\odot}}}
\newcommand{\sqas}{\Box ^{\prime\prime}}
\newcommand{\as}{ ^{\prime\prime}}
\newcommand{\am}{^{\prime}}
\newcommand{\bdm}{\begin{displaymath}}
\newcommand{\edm}{\end{displaymath}}
\newcommand{\beq}{\begin{equation}}
\newcommand{\eeq}{\end{equation}}
\newcommand{\bit}{\begin{itemize}}
\newcommand{\eit}{\end{itemize}}
\newcommand{\ben}{\begin{enumerate}}
\newcommand{\een}{\end{enumerate}}
\newcommand{\bfi}{\begin{figure}[htb]}
\newcommand{\bpfi}{\begin{figure}[p]}
\newcommand{\lbol}{$\rm L_{Bol}$}

\newcommand{\av}{\rm A_V}

\shorttitle{Mid-Infrared Imaging of Seyfert Nuclei}
\shortauthors{Krabbe, B\"oker, \& Maiolino}

\begin{document}


%%%%%%%%%%%%%%%
% Useful definitions during manuscript preparation
%%%%%%%%%%%%%%%

% Uncomment this to ignore figures during LaTeX compilation
% \def\epsfbox#1{}

% Macros to mark places where numbers remain to be added (\xx), where something
% remains to be done (\tbd), or where a comment is placed for the co-authors
% (\com).

\def\xx{$^{[xx]}$}

\def\tbd#1{{\baselineskip=9pt\medskip\hrule{\small\tt #1}
\smallskip\hrule\medskip}}

\def\com#1{{\baselineskip=9pt\medskip\hrule{\small\sl #1}
\smallskip\hrule\medskip}}

\title{N-Band Imaging of Seyfert Nuclei and the MIR-X-Ray Correlation}


\author{Alfred Krabbe}
\affil{University of California Berkeley, 366 Le Conte Hall,
Berkeley, CA 94720-7300}
\email{krabbe@ssl.berkeley.edu}

\author{Torsten B\"oker\altaffilmark{1}} \affil{Space Telescope
Science Institute, 3700 San Martin Drive, Baltimore, MD 21218}
\email{boeker@stsci.edu}

\and

\author{Roberto Maiolino}
\affil{Osservatorio Astrofisico di Arcetri, Largo E. Fermi 5,
	50125 Firenze, Italy}
\email{maiolino@arcetri.astro.it}


\altaffiltext{1}{Affiliated with the Astrophysics Division, Space
Science Department, European Space Agency}


%%%%%%%%%%%%%%%%%%%%%%%%%%%%%%%%%%%%%%%%%%%%%%%%%%%%%%%%%%%%%%%%%%%%%%
\begin{abstract}

We present new mid-infrared (N-band) images of a sample of eight nearby
Seyfert galaxies.  In all of our targets, we detect a central
unresolved source, which in some cases has been identified for the
first time. In particular, we have detected the mid-infrared emission
from
the active nucleus of NGC\,4945, which previously remained
undetected at any wavelength but hard X-rays. We also detect
circumnuclear extended emission in the Circinus galaxy along its major
axis, and find marginal evidence for extended circumnuclear emission
in NGC\,3281.

The high spatial resolution ($1.7\as$) of our data allows us to
separate the flux of the nuclear point sources from the extended
circumnuclear starburst (if present).  We complement our sample with
literature data for a number of non-active starburst galaxies, and
relate the nuclear N-band flux to published hard ($2-10\kev$) X-ray
fluxes. We find tight and well-separated correlations between
nuclear N-band flux and X-ray flux for both Seyfert and starburst
nuclei which span over 3 orders of magnitude in luminosity. We demonstrate
that these correlations can be used as a powerful classification
tool for galactic nuclei.

For example, we find strong evidence against NGC\,1808
currently harbouring an active Seyfert nucleus based on
its position in the mid-infrared-X-ray diagram.
On the other hand, we confirm that NGC\,4945 is in fact a
Seyfert~2 galaxy.
\end{abstract}
%%%%%%%%%%%%%%%%%%%%%%%%%%%%%%%%%%%%%%%%%%%%%%%%%%%%%%%%%%%%%%%%%%%%%%
%%%%%%%%%%%%%%%%%%%%%%%%%%%%%%%%%%%%%%%%%%%%%%%%%%%%%%%%%%%%%%%%%%%%%%
\section{Introduction}
The classification of ``active'' galaxies based on their optical
spectra alone provides an incomplete or even deceiving description of
the true nature ot this class of objects.  In some ``classical''
starburst galaxies, X-ray and VLBI observations have found indications
for the presence of an active galactic nucleus (AGN) \citep*{mas99,
vig99, smi98}. Mid-infrared (MIR) spectroscopy with ISO has shown
that some galaxies, which were optically classified as AGNs, are in fact
dominated by starbursts. Conversely, the presence of a heavily obscured AGN
was revealed in some systems previously classified as starburst
\citep*{gen98,lut99a,rig99}.

Mid-IR observations are particularly useful
for detecting dust-enshrouded AGNs for a number of reasons. Most
importantly, the MIR is much less affected by dust extinction
than, e.g., the optical regime
\cite[$A_{8-13\mum}\approx0.06\cdot \av$,][]{lut96}.
While the extinction is even lower at far-IR wavelengths,
warm dust emission in the  $60-100\mum$ regime
can be powered both by AGNs and starbursts and is therefore not
distinctive of the nature of activity. In contrast,
significant emission from hot dust at wavelengths between
2 and $10\mum$ is generally characteristic of heating by an
active nucleus \citep{oli95,mai98,moo96a,lut99a,cla00}.

Finally, the MIR allows diffraction-limited,
arcsecond-scale imaging from large ground-based telescopes, in
contrast
to observations at far-IR wavelengths which have to rely on space
observatories with limited aperture sizes and hence limited
resolution.
As a case in point, MIR spectroscopy with ISO is often unable to
detect the nuclear hot dust emission, since the host galaxy and
circumnuclear starburst emission dilute the nuclear AGN emission in
the large beam of the ISO observations.

A classical example is NGC\,4945. This galaxy hosts one
of the brightest AGNs at $100\kev$, although the X-ray source is heavily
absorbed \cite*[$\rm N_{H}\approx 5\cdot 10^{24}\>cm^{-2}$,][]{gua00}.
However, both optical and near-IR data do not show any indication for
the presence of the active nucleus and appear, instead, dominated by a
powerful nuclear starburst along with its superwind cavity
\citep{moo96b,mai99,mar00}. Even the MIR ISO spectra
appear dominated by the starburst \citep{spo00}.

Because of the reasons outlined above, sensitive MIR observations
with high spatial resolution hold the promise to
identify heavily obscured AGNs within starburst galaxies, even if
they are missed in optical and near-IR imaging and/or in
ISO spectroscopy, both because of heavy obscuration or dilution
  by the nuclear starbursting activity.
If such MIR observations were to find that the fraction of galaxies
with AGNs is significantly higher than estimated from
optical, near-IR, or ISO MIR surveys,
it would provide support to the idea of a
connection between starbursts and AGNs, which has been
suggested by various studies \citep{hec97,mai95}.
%%%%%%%%%%%%%%%%%%%%%%%%%%%%%%%%%%%%%%%%%%%%%%%%%%%%%%%%%%%%%%%%%%%%%%
\section{Observations and Data Reduction}
The observations described here were carried out with MANIAC, the
Mid- And NearInfrared Array Camera \citep{boe97}. The data were taken
during a four week period in February/March 1998
at the ESO/MPI $2.2\m$ telescope in LaSilla, Chile. Most galaxies were
observed during multiple nights with varying integration times.
Table~\ref{tbl-obs} summarizes the target list and observational
parameters.

All galaxies were observed through the MANIAC N-band filter which
comprises the spectral region between 8 and $13\mum$. We employed
standard chopping and beam switching
techniques to account for the high and rapidly variable sky emission
in the mid-infrared regime. In brief, a large number of frames with
typical integration times of a few milliseconds are co-added in
a hardware buffer for both the on-source and off-source
chopper position. Both buffer sums are saved to the MANIAC control
PC after typically 30~s, corrected for bad pixels,
and subtracted from each other. We offset the telescope
by exactly one chop throw about every minute to correct for the
slightly different illumination
patterns in the two chop beams. Because most of our objects
are fairly compact, we decided to keep the chop throw small enough
that the ``off'' beams still fell on the chip. Subtraction of the
two difference images images obtained at both beam positions then
yields a triple image of the source in which
the thermal background from sky and telescope
is reduced to about\footnote{The residual variations
are due to non-perfect centering of the chopping secondary. They
mainly cause some ambiguity in the zero level of the images, and
lead to increased uncertainty of the extended emission.}
$10^{-5}$. This procedure, which has been described
in more detail in \citet*{boe98}, improves the
signal to noise ratio for a given integration time, since the
``off'' beam images can be combined with the center image. However,
in the case of an extended source, overlap of the negative beams
with the center image can cause confusion. While there are algorithms
that can recover the information in the overlap regions \citep*{ber00}
we did not attempt to employ such reconstruction methods because
we are mainly interested in the nuclear, often unresolved, source for
which this problem is not important.

Since the data for many of our targets were taken over several nights
with varying seeing conditions, we smoothed the maps of the individual
nights to the worst nights seeing profile by convolving with
appropriate Gaussians.  The maps then were shifted to a common center
position, and median combined.  Images taken under the best seeing
conditions were used to determine the upper limits on the size of the
point-like nuclear source for each object (see Table~\ref{tbl-obs}).

In order to discriminate between the flux of the unresolved nuclear
source and the underlying extended emission, we constructed a mean
Point Spread Function (PSF) from standard star images close in time to
the respective galaxy observations during each night.  More
specifically, we weighted the standard star images by each night's
effective integration time on the galaxy before averaging.  We then
scaled the resulting ``galaxy specific'' PSF appropriately to subtract
the nuclear source, and derived the integrated flux.  The scale factor
was determined such that no residual unresolved emission remained
after subtraction while assuring a smooth interpolation of the
underlying extended emission (if present).  The quality of the
subtraction of the point source and remaining uncertainties were
assessed by inspecting the images.  Table~\ref{tbl-photometry} lists
the calibrated fluxes of both emission components of all target
galaxies together with their $1\sigma$ uncertainties. The
uncertainties include contibutions from the absolute calibration,
background level, and PSF fitting.

Flux calibration of the data was performed via observations of the
stars $\lambda$~Vel and $\gamma$~Cru, with assumed N-band magnitudes
of -1.78 and -3.36, respectively. We have adopted the Vega-based
photometric system, with the N-band zeropoint corresponding to $42.6\jy$.
%%%%%%%%%%%%%%%%%%%%%%%%%%%%%%%%%%%%%%%%%%%%%%%%%%%%%%%%%%%%%%%%%%%%%%
\section{Results}
\subsection{The MANIAC Sample}
The objects observed with MANIAC were selected
to have at least some indications of nuclear (Seyfert) activity and
non-negligible mid-IR emission. More specifically,
we selected observable ``active'' nuclei whose central
10$\mu$m emission is brighter than $\sim$0.3 Jy based on previous
ground-based observations \citep[$\sim 5\as$ aperture, e.g.][]{mai95}.

We present the resulting N-band maps of our sample of galaxies in
Figure~\ref{fig-maps}. Table~\ref{tbl-photometry} summarizes
the N-band photometry of the
unresolved nuclear component and of the extended emission (if present)
as derived from these images.
We also list in Table~\ref{tbl-photometry} literature values for the
visual extinction towards the central region
as inferred by various indicators such as narrow emission lines,
PAH features and near-IR colors of the stellar population which
typically trace the extinction on scales of a few hundred pc.
More details on our extinction estimates are given below in
the discussion of individual objects.
The adopted extinction values were used to derive the unobscured
N-band flux of the nuclear point source by assuming that it
is affected by the same reddening as the circumnuclear region. We
caution, however, that in the context of the unified model
for Seyfert nuclei, the $10\mum$ emission
is believed to originate from an obscuring torus on scales
much smaller than the circum-nuclear starburst, typically
a few pc \citep[e.g.][]{pie92}. Therefore, the
extinction values Table~\ref{tbl-photometry} more likely represent
lower limits for the true extinction towards to MIR emission.
This issue is discussed further in \S\ref{subsec:correlation}.

In addition, Table~\ref{tbl-photometry} contains X-ray fluxes
in the $(2-10)\kev$ energy band. These were taken from the literature,
and typically have been corrected for the photoelectric absorption as
measured in the same band. Details are again given in the discussion
of the individual objects below. The extinction inferred from
the optical and IR indicators is generally much lower than that
derived
from gas column densities as measured in the X-rays
by assuming a Galactic dust-to-gas ratio and extinction curve.
This effect is at least partly due to the larger size of the region
sampled by optical and IR indicators as discussed above (few $100\pc$)
compared to the nuclear X-ray source and the putative obscuring
torus. Also, various pieces of evidence
suggest that the dust in the circumnuclear
region of AGNs is characterized by properties significantly different
from the Galactic diffuse interstellar medium which is most
likely the reason for the reduced $\rm A_V/N_H$ ratio commonly
observed in this class of objects \citep{lao93,mai00a,mai00b}.

In what follows, we briefly discuss the N-band morphology as derived
from the maps in Figure~\ref{fig-maps}, the X-ray data, and extinction
estimate of all sample targets on an individual basis.
%---------------------------------------------------------------------
\subsubsection{NGC\,1808}
NGC\,1808 is a morphologically peculiar starburst galaxy
\citep{kra94}, which has being suspected of harbouring a hidden
Seyfert nucleus \citep{ver85,awa96,kot96}.  In particular, the decline
of the hard X-ray flux suggests that the nuclear activity has been
gradually fading during the past few years \citep{bas99}.  Our N-band
image shows a pronounced nucleus sourrounded by strong extended
emission. The extended emission can safely be attributed to the
extended starburst \citep{kra94}. The total N-band flux within $20\as$
is in excellent agreement with the $12\mum$ IRAS flux of $4.4\jy$.
The nuclear point source only accounts for less than 8\% of the total
flux.
%---------------------------------------------------------------------
        \subsubsection{NGC\,4945}
NGC\,4945 is a bright infrared galaxy revealing a powerful starburst
in its nuclear region \citep{ric88}.  Evidence for a heavily
obscured AGN comes from hard X-ray observations \citep{don96,gua00}.
However,
NGC\,4945 is peculiar in that the the AGN has not yet been succesfully
traced, neither in the NIR \citep{mar00} nor in the MIR \citep{spo00}.
So far, the role of the AGN for the bolometric luminosity has remained
unclear. Our N-band image reveals for the first time
an unresolved point source at the nucleus of NGC\,4945
together with strong extended emission along the major axis of the
galaxy. The extracted observed N-band flux of the central source is
only 8\% of the total N-band flux of NGC\,4945, which has been
determined within an ellipse $22\as\times8.7\as$ tilted to match the
major axis of the galaxy (Table ~\ref{tbl-photometry}).  Our total
N-band flux within this area is consistent with ISO observations
\citep{spo00}. The low value for the nuclear emission in conjunction
with the dilution of the much larger ISO beam explains why ISO has not
detected the AGN.

The extinction towards the central region can be assessed from MIR
data. \cite{spo00} and \cite{lut99b}
determined the spatially averaged ($\sim 20''$) extinction towards the
circumnuclear starburst and find $\av = 36$ within a 50\% error. This
value is probably a lower limit since usually the foreground
extinction rised towards the nucleus. Indeed, the nuclear ($3''$)
ISOCAM spectrum of NGC\,4945 indicates a much higher extinction,
based on the ratio of the 6.2$\mu$m and 7.7$\mu$m PAH features \citep{mai99}.
Here, we adopt $\rm A_V\approx 60$ with an uncertainty of 40\%.
%---------------------------------------------------------------------
\subsubsection{Circinus}
Circinus is the closest {\it bona fide} Seyfert~2 galaxy. It is
characterized
by a prominent ionization cone, extending to the North-West, and by
a circimnuclear starburst ring with a radius of about $10\as$, i.e.
$200\pc$ \citep{wil00, mar94}.
The active nucleus of Circinus is heavily obscured by a large
column density of gas \citep[$\rm N_H \approx 4\times
10^{24}\>cm^{-2}$,][]{mat99}.

The N-band map of Circinus is dominated by the central nuclear
point souce. Its flux is so high that limits in the bandwidths of the
electronic detector readout circuitry result in a notable elongated
shadow of the nucleus towards south. Since the emission
morphology can not be reliably determined over the affected
area, it has been masked out in the image.
The eastern and western boundaries of the image are defined
by residual signals from the negative beams due to chopping
and nodding, which have also been masked out.

Our image also shows circumnuclear N-band emission, which has not
been detected previously. This continuum emission is rather smothly
distributed but shows
a shallow peak with a flux level of $\rm 17\mjy/\sqas$ offset by
$(+5.5\as , +11\as)$ from the nucleus. This position coincides
with the intersection of the major axis and the circumnuclear
ring where a region of intense star formation is found
\citep{wil00, mar94}. Perpendicular to the major axis of
Circinus, in particular towards the north-west the region of the cone
of the NLR \citep{mar94}, we find significantly less extended
emission.

Subtracting a stellar PSF from the radial profile of the nuclear
source reveals a small deviation from the PSF profile, peaking at
about $5\mjy/\sqas$ at a radius of $3\as$ (Figure~\ref{fig-profile}).
Given the
limited spatial resolution of our images this possible extended
emission needs to be confirmed at higher spatial resolution. However,
it cannot be explained by seeing-induced broadening of the PSF.
%---------------------------------------------------------------------
\subsubsection{Centaurus A}
At a distance of $3.5\mpc$, Centaurus A is one of the closest AGNs in
our sample, along with NGC4945 and Circinus. The presence of a hidden
active nucleus is revealed both through its hard X-ray emission and
the presence of a radio jet \citep*{cla86}.
The notion of an AGN in Centaurus~A is also supported by our N-band
image which only reveals a prominent unresolved central source (not
shown in Figure~\ref{fig-maps}).
While \cite{mir99} found extended MIR emission from ISO observations, our
high resolution image is not sensitive to such diffuse emission on scales
much larger than the MANIAC field of view. Nevertheless, the photometry of
the nuclear source is unaffected from this uncertainty because it is
measured differentially against the background.
%---------------------------------------------------------------------
\subsubsection{NGC\,6240}
Our N-band image of NGC\,6240 was obtained
during an earlier observing run in 1997.
It is dominated by the two nuclei of the well known
merger system \cite[e.g.][]{fri83} which are just
resolved in our map. The southern nucleus is about twice as bright
as the northern one (Table~\ref{tbl-photometry}).

The low-level structure in our map does indicate the presence
of faint extended emission in the vicinity of the nuclei,
consistent with \cite{ket97}. However, its contribution to the
overall flux is only minor.
For flux calibration, we used the photometry of \cite{ket97}, since
the MANIAC calibration was not yet well established for these earlier
observations.

Hard X-ray observations provide evidence for the presence of
a powerful AGN which is heavily absorbed by a column density of
$\rm N_H \approx 2\times 10^{24}\>cm^{-2}$ \citep{vig99}.
Their absorption-corrected $(2-10)\kev$ luminosity
is $\rm 3.6\times 10^{44}\>ergs/s$ ($\rm H_o = 50\kms /Mpc$)
which places this object in the QSO luminosity range
\citep{vig99,ree00}.
%---------------------------------------------------------------------
        \subsubsection{NGC\,3081 and NGC\,3281}
The N-band map of NGC\,3081 reveals a point source which is most
likely due to dust emission associated with the Seyfert~2 nucleus.
Any extended emission is below our detection limit, even if
integrated over an aperture of 25$\as$ in diameter
(Table~\ref{tbl-photometry}). Assuming that
the foreground extinction is comparable to values determined by
\citet*{goo94} from NIR lines towards the narrow line region, we
adopt $\av\,=\,4$.

The image of NGC\,3281 is also dominated by a
point source. In addition, there is marginal evidence for an extended
emission component in our map. While its detailed morphology is probably
not reliable, the residual flux within a 25$\as$ aperture after
subtraction of the unresolved component is about 13\% of the point
source flux. Despite the
uncertainties in the zero level of the map, this is significant.

If true, the integrated extended emission puts an upper limit to the
extended flux per beam to about $0.4\mjy$, which is more than
1000 times fainter than the nucleus. Our N-band flux density is
consistent with J through M photometry by \citet{sim98}.
%---------------------------------------------------------------------
        \subsubsection{NGC\,1566}
NGC\,1566, the brightest member of the Doradus group, is an almost
face on spiral galaxy with a Seyfert nucleus. The exact
classification, however, varies in the literature between Seyfert~1
\citep*[e.g.][]{gre92} and Seyfert 2 \cite[e.g.][]{all85}. Our N-band
map (not shown in Figure~\ref{fig-maps}) reveals a point source with 
no significant circumnuclear emission above our sensitivity limit.

For NGC\,1566, no $(2-10)\kev$ data are available.
 From Einstein data, \cite{gre92} determine a $(0.5-4.5)\kev$ flux
of $\rm 6\times 10^{-12}\ergs \cm$ (converted to our H$_0$).
\cite{ehl96} do not find evidence for foreground absorption in excess to
the Galactic value, based on ROSAT data. By extrapolating the
$(0.5-4.5)\kev$ flux
to the $(2-10)\kev$ band with a typical photon index of 1.7, we obtain
$\rm F_{2-10\kev}= 6^{+3}_{-2}\times 10^{-12}\ergs \cm$, where the large errors
account for the uncertainty in the spectral slope.
%---------------------------------------------------------------------
\subsection{Complementary Literature sample}
In order to compare the MIR and X-ray properties of Seyfert galaxies
with those of prototypical starburst galaxies,
we searched the literature for additional objects with
both high resolution MIR images and hard X-ray fluxes.
Table~\ref{tbl-comp} lists the results of our search including
the references that contain the relevant data.
Some of the objects (NGC\,7552 and NGC\,253)
have also been observed with the MANIAC camera, and have been reduced in
an identical fashion. In general, there is a scarcity of high-resolution
MIR maps of starburst galaxies, a fact that
is likely to change in the
near future due to the comissioning of a number of large IR telescopes
such as Gemini or the Large Binocular Telescope. Specific aspects of the
objects in Table~\ref{tbl-comp} are discussed in the appendix.
%%%%%%%%%%%%%%%%%%%%%%%%%%%%%%%%%%%%%%%%%%%%%%%%%%%%%%%%%%%%%%%%%%%%%
\section{Interpretation and Discussion}
%---------------------------------------------------------------------
\subsection{The nuclear 10$\mum$ emission}
Generally,
the mid-IR continuum emission from the central region of AGNs is ascribed to
thermal emission from dust which is irradiated by the nuclear UV-optical
radiation, and possibly associated to the obscuring torus postulated by the
unified model \citep*{pie92,gra97,lao93}. Typical dust temperatures are
about 300 K. For the luminosities of most of the Seyfert nuclei considered
here (\lbol\, of a few times $10^{10}\lsun$) an equilibrium
temperature of about $300\>K$ is reached at a distance of a few pc from
the nuclear source \citep{lao93}.
With our high signal-to-noise mid-IR images we can set an upper
limit\footnote{Based on the comparison with stellar
PSFs and taking into account fitting uncertainties.}
of about $0.5\as-0.65\as$ on the size of the nuclear unresolved source,
which corresponds to a projected size
of $\approx10\pc$ (for Circinus, Cen\,A, NGC\,4945),
$\approx40\pc$ (for NGC\,1808, NGC\,1566), and between $90\pc$ and
$260\pc$ for the more distant targets, i.e. consistent with the size
of the warm dust emitting region discussed above.

Compared to normal and SB galaxies, AGNs are generally very effective in
heating dust to temperatures high enough to emit at mid-IR wavelengths
\citep{gen98,lut99a}. As a consequence, the detection of an unresolved
mid-IR source in the galactic nucleus is generally a good indication
for the presence of an AGN. In this regard, the identification of nuclear
unresolved MIR emission in our N-band maps of NGC\,4945, NGC\,3081,
NGC\,3281, and NGC\,1566 is a case in point. In particular, all previous
attempts to detect the nucleus of NGC\,4945 have remained unsuccessful,
even at NIR wavelengths \citep{spo00,mar00}.

On the other hand, the presence of an unresolved MIR nucleus alone
is not necessarily a proof for the presence of
an AGN. Starburst galaxies may also harbor
unresolved nuclear emission. For example, both NGC\,7552 and NGC\,1808
have nuclear point sources in the MIR, yet they show no signs for an AGN
(see discussion below).

In summary, detecting and isolating the nuclear $10\mum$ flux provides a
necessary prerequisite to further constrain the nuclear activity of
galaxies, but it alone is not sufficient to distinguish between various
kinds of nuclear activity.
%---------------------------------------------------------------------
	\subsection{X-ray emission}
The X-ray emission from AGNs is generally assigned to the nuclear
engine itself. The strong UV/optical continuum radiation is believed
to be responsible for heating the circumnuclear dust, which in turn
is the source of the nuclear MIR emission via thermal radiation.
Therefore, a correlation between X-ray and
MIR nuclear radiation in AGNs would be a natural result.

In most type 2 Seyferts, large column densities of cold gas along the
line of sight absorb the soft X-ray flux. In particular, the superposition of
the absorption edges of various metals introduces a photoelectric cutoff
towards the lower energies. The energy of the cutoff increases with the gaseous
column density along the line of sight.
However, as mentioned above,
at energies above the photoelectric cutoff (generally in the range $2-10\kev$),
it is possible to observe the transmitted radiation and to infer the intrinsic
(absorption-corrected) luminosity of the nuclear source \citep{bas99}.

If the absorbing
column is thick to Compton scattering ($\rm N_H > 10^{24}\>cm^{-2}$),
the radiation is completely absorbed up to at least $10\kev$.
Therefore, the traditional ``hard'' X-ray band between 2 and $10\kev$
is dominated by scattered and reflected radiation.
If the column density is
not higher than about $\rm 10^{25}\>cm^{-2}$, X-rays in the $10-100\kev$
band are at least partly transmitted, and can indeed be observed
with X-ray satellites sensitive in this band (e.g. BeppoSAX).
In such cases, the intrinsic X-ray luminosity of the nuclear
source can be inferred. Examples for this method include
NGC\,4945, Circinus, or NGC\,6240 \citep{gua00,mat99,mat00,vig99}.

In contrast, the $(2-10)\kev$ emission of
starburst galaxies without an AGN is mainly produced by
young supernova remnants, X-ray binaries and/or hot diffuse gas,
the latter possibly produced in supernova induced
superwinds \citep{bev00,cap99}.
However, the cumulative hard X-ray emission of
these components is significantly weaker than the emission produced
by AGNs \cite*[e.g.][]{dav92}. On the other hand, the young stellar
population in starburst galaxies is quite effective in heating
the dust which radiates MIR thermal emission and in exciting the PAH
features which also contributes significantly to the MIR observed flux.

Therefore, a comparison of the nuclear $10\mum$ emission with the intrinsic
(i.e. absorption-corrected) nuclear $(2-10)\kev$ X-ray flux should
allow to discriminate between star formation and active nuclei.
In practice, however, one has to account for an additional
complication. The nuclear component of the $(2-10)\kev$ radiation
cannot be determined with the same spatial resolution as the
nuclear $10\mum$ emission, because the angular resolution of satellites with
hard X-ray sensitivity such as ASCA or BeppoSAX is limited to $\approx 1\am$,
so that in cases where both starburst and AGN activity is present, the
X-ray measurements may contain contributions from both sources.
For his reason one has to be cautious when determining the nature
of these nuclei based on $(2-10)\kev$ data alone.
Only the combination of nuclear MIR and hard X-ray emission seems
to provide a reliable tool for constraining the
nature of the nuclear activity of theses sources, as demonstrated in
the next section.
%---------------------------------------------------------------------
	\subsection{MIR--X-ray Correlation}\label{subsec:correlation}
In Figure~\ref{fig-plot}, we compare the $10\mum$ flux density of the
unresolved nuclear component to the observed X-ray fluxes in the
$(2-10)\kev$ band after correction for their respective absorptions.
The error bars account for uncertainties in the photometry itself and
in the estimates for foreground extinction (see
Table~\ref{tbl-photometry}).  Figure~\ref{fig-plot} shows that, within
the statistical limits of our sample, the nuclear MIR flux density of
type 2 Seyferts is proportional to the intrinsic hard X-ray flux.
Four other Seyfert~1 galaxies, for which we found comparable data
(NGC\,7469, NGC\,4051, NGC\,4151 and NGC\,3227), follow the same
correlation, which is robust against modifying single values by
factors of the order of 2.  A similar correlation holds for the
starburst galaxies, however, their hard X-ray flux is about a factor
of 15 less than for Seyferts.  The straight lines in
Figure~\ref{fig-plot} are linear fits to the data which have been
forced to cross the origin.  NGC\,6240 lies significantly
above the correlation and will be discussed further down.

The proportionality between the N-band and X-ray flux densities in
Figure~\ref{fig-plot} is encouraging.  The correction on the X-ray
emission is quite accurate: even in the most extremely obscured cases
(NGC\,4945, NGC\,6240) the uncertainty is below a factor of 2, which
is small when reported on a logarithmic scale.  In previous
attempts to find similar correlations between X-ray and MIR/FIR fluxes
\citep[e.g.][]{san89,dav92}, no correlation had been found 
or the scatter was much larger, mostly due to lack
of spatial resolution and sensitivity.  Having established the
existence of a correlation in flux, we can now convert to luminosities 
in order to investigate correlations between intrinsic properties of the
galaxies.  By doing so, i.e. multiplying the MIR and X-ray fluxes of each 
object in both axis by the square of the distance and some factors, the 
data points in Figure~\ref{fig-plot} will be rearranged parallel to the 
solid line, thus conserving the scatter of the correlation.  The result is
displayed in Figure~\ref{fig-lum}.  Within the uncertainties and
excluding NGC\,6240 for the moment, the distribution is still
compatible with a slope of 1 (solid line in Figure~\ref{fig-plot}),
covering more that 3 orders of magnitude in luminosity.

The location of NGC\,6240 in Figure~\ref{fig-lum} is
above the correlation for the other Seyfert galaxies.  More
specifically, is is overluminous in hard X-rays by about an
order of magnitude.  Possibly, the dust extinction towards the nucleus
of NGC\,6240 is much higher than assumed based on NIR
colors \citep{tec00,ale00}.  Indeed, NGC\,6240 is the most distant object of
our sample and the observations sample larger spatial scales
($\sim$kpc).  It is therefore likely that extinction estimates are
diluted by regions farther from the nucleus with lower optical depths
and that the intrinsic mid-IR emission of NGC6240 is higher than
derived from our data. Thus, a much larger fraction of the radiation of the
very hot dust may already have been reprocessed and shifted to longer
wavelengths before it left the nuclear environment.  A foreground
extinction in the range A$_{\rm V}\sim 50$ mag is required to
increase the dereddened nuclear N-band emission such that
NGC\,6240 complies with the correlation of the other
AGNs.

The nuclear $10\mum$ fluxes were corrected for the extinction observed
in the circumnuclear region as inferred from either NIR color maps,
recombination line ratios or from the ratio between PAH features.  The
fact that the tight correlation between $10\mum$ and X-ray fluxes
appears to be the same for both Seyfert types indicates that the
$10\mum$ emission from the nuclear source is not affected by an
extinction much higher than inferred for the circumnuclear region.
This indicates that either the $10\mum$ emitting region is more
extended than the putative obscuring torus predicted by the unified
model for AGNs, or that the optical depth of the torus is not much
higher than that of the circumnuclear region for these objects.  This
result appears to be in contradiction with previous studies which
claimed significant differences in the nuclear $10\mum$ emission
between obscured and unobscured Seyferts \citep{hec95, mai95, giu95}.
This discrepancy can probably be explained by limited angular
resolution and sensitivity of those earlier studies.

An alternative explanation which is likely to important at least to some
extent stems from the selection criteria of our sample. Because the
MANIAC targets were selected to have strong nuclear MIR emission, the
sample is potentially biased towrads AGNs which are little absorbed at
this wavelength. Therefore, possible differences between different
classes of Seyferts in terms of their mid-IR absorption might have been missed.
Additional high resolution MIR observations of Seyfert
nuclei are required to solve this issue.

The X-ray emission from the comparison sample of pure (to present
knowledge) starburst nuclei in Figure~\ref{fig-lum} is significantly 
lower and appears to follow a line with a slope different from
proportionality.  We caution that the apparently different slope 
might be caused by the small number statistics of our sample. On the 
other hand, it can plausibly be explained by the very different aperture
sizes that were used for the measurements: about $2\as$ for the N-band
and between $30\as$ and about $1\am$ for the X-rays.  Thus, for 
starburst galaxies, one would measure hard X-rays from almost the entire
extended starburst region, but only a distance-dependent fraction of
the N-band emission. In support of this argument, we note that using 
a larger aperture for the MIR measurement, e.g. the $12.5\mum$ IRAS 
luminosity from Table~\ref{tbl-comp}, the proportionality slope appears
to be restored (although at increased scatter), as shown in 
Figure~\ref{fig-lum} on the lower right.  The vertical distance
between the two proportionality lines is a factor of 190, which
indicates the different physical processes that prevail in these two
classes of objects.  More specifically, classical starburst galaxies
appear to have X-ray fluxes that are a large factor of almost
200 below those of AGNs with similar $10\mum$ emission.

Yet, we cannot completely rule out that the nuclear unresolved
$10\mum$ source in some of the starburst nuclei (e.g. NGC~7552) is the
counterpart of a heavily obscured AGN. Indeed, if the nuclear
obscuring medium is completely Compton thick ($\rm N_H >
10^{25}\>cm^{-2}$), the $(2-10)\kev$ emission from the AGN would be
completely suppressed and instead be dominated by the starburst
radiation.

An important result of our study is that in Figure~\ref{fig-plot},
NGC\,4945 definitely falls into the region occupied by AGNs, despite
the rather large uncertainty of its intrinsic $10\mum$ flux.  This
result suggests that the unresolved nuclear MIR source is indeed
powered by the obscured active nucleus detected in hard X-rays.

The position of NGC\,1808 in the regime of starburst galaxies suggests
that the active nucleus in this galaxy - if present at all -
has faded \citep{awa96,bas99}. Currently, the X-ray and MIR emission
are instead dominated by an (unresolved) circum-nuclear starburst.

It is also interesting that the archetypical starburst galaxy M\,82
falls into the regime of the Seyfert nuclei.  This statement is robust
against any uncertainty in the X-ray flux due to the observed
variability.  In order for M\,82 to fall into the regime occupied by
starburst nuclei, its X-ray flux should have to be a factor of 30
lower in Figure~\ref{fig-plot}, certainly incompatible with
observations.  We therefore concluded that M\,82 is likely to harbor
an AGN, in agreement with \cite{pta99} and \cite{mas99}.  Meanwhile,
subsequent observations with Chandra at about $1\as$ spatial
resolution suggest that a highly variable source about $9\as$ off the
nucleus, which is dominating the hard X-ray flux, is compatible with a
mid-mass Black Hole candidate \citep{kaa00, mas00}.  If this can be
proven it would explain the location of M82 in Figure~\ref{fig-plot},
since the same kind of physics is applied.  In Figure~\ref{fig-lum},
the N-band luminosity of the M82 Black Hole candidate has been
corrected for the $9\as$ offset and slightly shifted its position.

It is important to note that the position of M82 in the diagram
largely depends on the MIR aperture.  This provides additional support for
the argument already discussed with NGC\,4945: a weak AGN can
easily be diluted and ``drowned'' in a large beam due to circumnuclear
starburst activity.  This demonstrates the need for observations with
the highest possible spatial resolution if the correlation is to be extended
towards more distant targets.  Otherwise, less luminous AGNs will
be easily missed at higher distances.
%---------------------------------------------------------------------
	\subsection{Extended MIR emission}
Previous studies found evidence for extended mid-IR emission in Seyfert 2
galaxies based on the comparison between ground-based
observations and IRAS data \citep{mai95}. Such extended emission is ascribed
to enhanced star forming activity in the host galaxies of these obscured active
nuclei. Our images show such extended emission in most of the
objects of our sample, thus supporting and reinforcing those early results,
although with limited statistics. Furthermore, our images indicate that
generally the starforming activity occurs in the vicinity of the active nucleus
(within a few 100 pc). This supports the idea of a tight connection between
starburst and AGN activity as
claimed in the past by various authors based both on observational and
theoretical grounds \citep*{hec89,gon98,oli99,rod87,nor88,col99}.
%---------------------------------------------------------------------
\section{Conclusions}
We have presented new N-band maps of a sample of 8 nearby Seyfert galaxies.
In all of our targets, we have identified an unresolved nuclear emission
source, some of which were previously undetected.
In those cases where additional extended emission is present,
we have separated the unresolved component by means of PSF fitting.
This demonstrates the importance of high resolution MIR imaging for
detecting AGN and isolating them from the extended emission component.

After complementing our sample by literature data for a number of
starburst galaxies and some additional Seyferts, we have compared the MIR
luminosity of the nuclear sources to their hard X-ray emission.
We find that both ``pure'' starbursts and AGN follow a tight MIR--X-ray
correlation, but the two populations are separated by about a factor of
15 without and 190 with aperture correction
in their hard X-ray fluxes. We propose to use this correlation as a
powerful diagnostic tool to distinguish between these two classes
of objects in cases where evidence for nuclear activity is ambiguous.
%%%%%%%%%%%%%%%%%%%%%%%%%%%%%%%%%%%%%%%%%%%%%%%%%%%%%%%%%%%%%%%%%%%%%%%%%%%%%
\acknowledgements We are grateful to a number of people who have
contributed significantly to the success of the MANIAC observing
campaigns.  In particular, we are indebted to John Storey, Craig
Smith, Thomas Lehmann, and Gerd Jakob.  We also wish to thank the ESO
staff at LaSilla for their support, and especially Ueli Weilenmann and
Hans Gemperlein for help with the chopper control electronics.  We
thank the referee, Bob Joseph, for helpful comments to improve this
paper.
%---------------------------------------------------------------------
         \section{Appendix}
The nuclear emission of the starburst galaxy {\bf NGC\,253} was derived from
the $12.8\mum$ continuum image of \cite{boe98} which is in good
agreement with that published by \cite{ket99}. We estimate the
background level at the location of the nucleus of NGC\,253 to be of the
order of 20\% of the peak flux in Figure~2 of \cite{boe98}, whereas
the nucleus is at 40\% of the peak flux in the figure. Accounting for 
the fact that
the spectral band of the data of \cite{boe98} is not identical with
the N-band, we estimate the N-band flux to be $(400\pm100)\mjy$.

The nuclear emission of the prototypical starburst galaxy {\bf M\,82} was
derived using the data of \cite{tel91b}. From their Figure~5, we
estimate that the upper limit to the contribution of the nucleus is $200\mjy$,
which corresponds to one contour spacing in their figure. The foreground
extinction was estimated using \cite{tel91b} and \cite{sat95}. M\,82
shows variable hard X-ray flux \citep{pta99,mas99}, which led these
authors to speculate about wether M\,82 might be harboring an AGN. We used
the amplitude of the variability \citep{pta99} as a realistic value
for the uncertainty of the X-ray flux level.

The Seyfert~1 galaxy {\bf NGC\,7469} was included in this list to compare
with the Seyfert~2 targets.
Taking into accout the differences in bandwidth and wavelength between
\cite{car88} and \cite*{mil94} we obtain an N-band
flux of the nucleus alone of $620\pm80\mjy$.

{\bf NGC\,7552} is a starburst galaxy with a prototypical circum-nuclear
starburst ring seen in the near- and mid-infrared. The N-band map
of \cite{sch97} also shows a faint nuclear source.
We obtained the N-band flux of the nucleus from Table~2A in
\cite{sch97}, estimating that about 1/3 of the flux within $2\as$ are
due to the background emission.

\cite{cha82} determined an absorption corrected $(0.2-4.0)\kev$ X-ray
flux of $(7.0\pm1.5)\cdot 10^{-12}\ergs\cm$, consistent with
\cite{mac81}, who find variable fluxes in the range
$(0.3-1)\cdot10^{-12}\ergs\cm$ .
They assumed a low absorbing column density consistent
with the derived low $\av$ number above. Assuming a shape of the X-ray
spectrum of NGC\,7552 similar to other starburst galaxies in \cite{pta99},
one can assess the $2-10\kev$ flux to be about 1/3 of the
$(0.2-4.0)\kev$ flux or $(2.0\pm2)\cdot10^{-13}\ergs\cm$.

{\bf NGC\,3690} is a merging system with an eastern and western
component. The sum of their nuclear N-band fluxes can be obtained
from \cite{ket97}.
The $(2-10)\kev$ hard X-ray flux is
integrated over both nuclei, with only low foreground
absorbing column density \citep{ris00,zez98}.

Individual hard X-ray fluxes for the two
dominant components of NGC\,3690 are not yet available.
However, the combined mid-infrared and
hard X-ray fluxes of both nuclei can safely be used in
this investigation since both nuclei show only starburst activity,
without any evidence for an AGN.

The nuclear N-band flux of the starburst galaxy {\bf NGC\,3256} was 
determined to be
1.7 Jy within $15\as$ \citep{gra84} and 1.9 Jy
within 14$\as\times 20\as$ \citep{rig99}. However, according to
the map in \cite{boe97} the nucleus is dominated by a point source
sourrounded by notable extended emission within this aperture. We
therefore adopt for the flux density
of the nucleus alone 0.9 Jy, which is about half of the value above.

{\bf NGC\,6946}:The background subtracted nuclear N-band flux density 
derived from the
$4\as$ resolution map of \cite{tel93} is 250$\pm50\mjy$,
assuming that the lowest contour reflects the background. N-band
aperture photometry of different sizes by \cite{rie76} sums to
$360\pm50\mjy$ within $4.3\as$. Therefore we average the N-band flux
density to $300\pm100\mjy$.

The absorption corrected $(2-10)\kev$ X-ray flux can be read off Figure~2
and Table~12 (source S3) of \cite{pta99a}.  However, comparison of
their data with ROSAT $(0.2-2.5)\kev$ data of \cite{sch94}, in
particular his Figure~4, clearly shows that the X-ray sources do not
necessarily coincide with the nucleus of NGC\,6946.  The offset
between S3 in \cite{sch94} and the nucleus is about $20\as$, which is
supported by comparing his optical plate with a high-resolution HST
image, e.g. by \cite{boe99}. Significant contribution to the
$(2-10)\kev$ flux from strong circumnuclear sources are probable and we
therefore only account half of the adopted $(2-10)\kev$ flux of
$(2.7\pm0.5)\cdot 10^{-12}\ergs\cm$ to the nucleus.

{\bf NGC\,4051}, {\bf NGC\,3227} and {\bf NGC\,4151}: Mid-infrared data
for these Seyfert 1.2 to 1.5 galaxies are obtained from \cite{dev87} and
\cite{leb79}, who used 6 and $8\as$ apertures, respectively.
For NGC\,4051 we used the X-ray flux prior to the temporary
fading of this source reported by
\cite{gua98}.  The errors on the X-ray fluxes account for the
variability of these sources \citep{yaq91,geo98b,gua98}.

%%%%%%%%%%%%%%%%%%%%%%%%%%%%%%%%%%%%%%%%%%%%%%%%%%%%%%%%%%%%%%%%%%%%%

%%%%%%%%%%%%%%%%%%%%%%%%%%%%%%%%%%%%%%%%%%%%%%%%%%%%%%%%%%%%%%%%%%%%%
%%%%%%%%%%%%%%%%%%%%%%%%%%%%%%%%%%%%%%%%%%%%%%%%%%%%%%%%%%%%%%%%%%%%%
%% No more than seven \figcaption commands are allowed per page,
%% so if you have more than seven captions, insert a \clearpage

\newpage
\figcaption[fig1c.eps]{
    \label{fig-maps}
    $10.5\mum$ N-band images of Seyfert galaxies. For all maps, north
    is up and east to the left. All maps are convolved to $1.7\as$
    resolution, and centered on the apparent
    $10\mum$ nucleus. Some of the images show
    residuals of the negative beam, e.g. NGC\,4945 or NGC\,3281.
    See text for more comments on individual galaxies.  }

\figcaption[fig2.eps]{
    \label{fig-profile}
    Radial profile of the Circinus $10\mum$ emission peak (thick line),
    compared to a PSF profile taken under similar atmospheric conditions.
    The PSF was rescaled to match the Circinus peak. The difference between
    both curves is shown on a 10 times magnified scale (right y-axis).}

\figcaption[fig3.eps]{
    \label{fig-plot}
    $10.5 \mum$ - $(2-10)\kev$ hard X-ray correlation of isolated 
Seyfert nuclei.
    The lines represent a fit of slope 1 (in log space) through both the Seyfert
    (excluding NGC\,6240) and the starburst sample. Both lines
    are a factor of 15 apart in X-ray flux.   }

\figcaption[fig4.eps]{
    \label{fig-lum}
    The same dataset as in Figure~\ref{fig-plot}, but converted to
    $(2-10)\kev$ and N-band luminosities.  Within the errors, the
    distribution is still compatible with a slope of 1.  The tight
    correlation suggests a physical connection between the $10.5\mum$
    and $2-5\kev$ luminosity, which seems independent of the Seyfert
    type.  The starburst galaxies follow a correlation with a different
    slope, which is, however, due to the different aperture between
    N-band and X-ray observations. Correcting for the aperture by
    using $12\mum$ IRAS fluxes (Table\ref{tbl-comp}) instead recovers
    a slope of 1 (rightmost correlation). See text for details.}

\newpage
\begin{deluxetable}{lcccccccc}
\tabletypesize{\scriptsize}
\tablecaption{Summary of Observations \label{tbl-obs}}
\tablewidth{0pt}
\tablehead{
\colhead{(1)} & \colhead{(2)} & \colhead{(3)} & \colhead{(4)} &
\colhead{(5)} & \colhead{(6)} & \colhead{(7)} & \colhead{(8)} &
\colhead{(9)} \\
\colhead{Galaxy} & \colhead{R.A}   & \colhead{Dec.}      &
\colhead{Distance} &
\colhead{Scale}  & \colhead{Type}  & \colhead{Obs. date} &
\colhead{$t_{int}$}  & \colhead{Nucl. FWHM}  \\
      		 & \colhead{(J2000)} & \colhead{(J2000)} & \colhead{[Mpc]} &
\colhead{[pc/$^{\prime\prime}$]} &   & 			 &
\colhead{[min]} & \colhead{[$^{\prime\prime}$]}
}
\startdata
NGC\,1566 & 04 20 00.6 & -54 56 17   & 17.6 	    & 85   & Sy1 &
02-98 & 24 & 1.1 \\
NGC\,1808 & 05 07 42.3 & -37 30 46   & 11.1 	    & 54   & Sy2 &
02-98 & 56 & 1.1 \\
NGC\,3081 & 09 59 29.5 & -22 49 35   & 28.9 	    & 140  & Sy2 &
02-98 & 125& 1.3 \\
NGC\,3281 & 10 31 52.0 & -34 51 13   & 43.0 	    & 209  & Sy2 &
02-98 & 48 & 1.0 \\
NGC\,4945 & 13 05 27.5 & -49 28 06   & $3.9^{(a)}$  & 18.9 & Sy2 &
02-98 & 83 & 1.2 \\
Cen\,A     & 13 25 27.6 & -43 01 09   & $3.5^{(b)}$  & 17.0 & Sy2 &
07-97 & 10 & 1.1 \\
Circinus  & 14 13 09.3 & -65 20 21   & $4.0^{(c)}$  & 19.4 & Sy2 &
02-98 & 54 & 1.1 \\
NGC\,6240 & 16 52 58.9 & +02 24 03   & $97.0^{(d)}$ & 471  & Sy2 &
07-97 & ?? & 1.1
\enddata

\tablecomments{(1-3): Galaxy name and coordinates. (4): Galaxy distance.
Unless otherwise noted, the values were derived from the galacto-centric
radial velocity from \cite{vau91}, assuming $\rm H_0=75\kms$. (6):
activity class (8): total on-source integration time (9):
FWHM of the nuclei from those sets of data obtained under the best atmospheric
conditions. Numbers are consistent with those of standard stars
obtained shortly after or before.}

\tablerefs{
(a) \cite{ber92}
(b) \cite{mir99}
(c) \cite{sie97}
(d) \cite{tac99}
}
\end{deluxetable}
%%%%%%%%%%%%%%%%%%%%%%%%%%%%%%%%%%%%%%%%%%%%%%%%%%%%%%%%%%%%%%%%%%%%%%%%
\clearpage
\begin{deluxetable}{lccccc}
\tabletypesize{\scriptsize}
\tablecaption{Photometry of MANIAC Sample  \label{tbl-photometry}}
\tablewidth{0pt}
\tablehead{
\colhead{(1)} & \colhead{(2)} & \colhead{(3)} &
\colhead{(4)} & \colhead{(5)} & \colhead{(6)} \\
\colhead{Galaxy} & \colhead{Nucleus} &\colhead{Extended} &
\colhead{$A_V$} & \colhead{Nucleus (corr.)} & \colhead{X-ray} \\
      & \colhead{[mJy]} & \colhead{[mJy]} & \colhead{[mags]} &
      \colhead{[mJy]} & \colhead{[$\rm 10^{-14}\>W\>m^{-2}$]}
}
\startdata
NGC\,1566 & $97\pm 15$      & - 	     & $2\pm 50\%^a$    &
$110^{+17}_{-15}$	& ${0.6^{+0.3}_{-0.2}}^{~j}$ \\
NGC\,1808 & $331\pm 30$     & $3900\pm 600$  & $5\pm 40\%^b$    &
$440^{+70}_{-60}$	& $0.11^{+0.06}_{-0.04}$ \\
NGC\,3081 & $136\pm 15$     & - 	     & $4\pm 25\%^c$    &
$170^{+21}_{-19}$ 	& $0.7^{+0.7}_{-0.3}$ \\
NGC\,3281 & $580\pm 30$     & $74\pm 57$     & $5+ 25\%^d$    &
$760^{+70}_{-40}$ 	& $2.5^{+0.6}_{-0.5}$ \\
NGC\,4945 & $105\pm 15$     & $1290\pm 140$  & $60\pm 40\%^e$   &
$2880^{+7900}_{-2100}$ & ${25^{+10}_{-7}}^{~k}$ \\
NGC\,5128 (Cen A) & $750\pm 55$     & - 	     & $30\pm 10\%^f$   &
$3900^{+760}_{-640}$ & $20^{+20}_{-10}$ \\
NGC\,6240 (north) & $67\pm 30$   & - 	     & $1.6\pm 30\%^g$  &
$70^{+33}_{-23}$ 	& ${14^{+7}_{-4.5}}^{~l}$ \\
NGC\,6240 (south) & $143\pm 70$   & - 	     & $5.8\pm 30\%^g$  &
$200^{+100}_{-70}$ 	& ${14^{+7}_{-4.5}}^{~l}$ \\
Circinus & $10590\pm 1050$ & $5920\pm 1100^m$  &
${10^{+30\%}_{-0\%}}^i$     & $21500^{+5500}_{-2000}$&
$45^{+56}_{-25}$
\enddata

\tablecomments{(1): Galaxy name (2): N-band flux of the unresolved
nuclear component
(3): integrated N-band flux of the extended emission component (4):
estimated extinction towards the
$10\mum$ emitting region (see discussion in text) (5):
extinction-corrected N-band flux of the
unresolved nuclear source (6): absorption-corrected $(2-10)\kev$
X-ray flux, as taken from
\cite{bas99} unless otherwise noted.}

\tablerefs{
(a) \cite{ehl96}; (b) \cite{kra94}; (c) \cite{goo94}; (d)
\cite{sim98}, lower limit;
(e) \cite{spo00}; (f) \cite{gen98}; (g) \cite{tec00}; (h) \cite{sch97};
(i) \cite{mai98}, lower limit; (j) \cite{ehl96}, \cite{gre92}, \cite{elv90};
(k) \cite{gua00}; (l) \cite{vig99}, value includes both nuclei;
(m) 1400 of which accounts for the extended north-western emission}
\end{deluxetable}

%%%%%%%%%%%%%%%%%%%%%%%%%%%%%%%%%%%%%%%%%%%%%%%%%%%%%%%%%%%%%%%%%%%%%%%%
%%%%%%%%%%%%%%%%%%%%%%%%%%%%%%%%%%%%%%%%%%%%%%%%%%%%%%%%%%%%%%%%%%%%%%%%
\clearpage
\begin{deluxetable}{lccccc}
\tabletypesize{\scriptsize}
\tablecaption{Photometry of Literature Sample  \label{tbl-comp}}
\tablewidth{0pt}
\tablehead{
\colhead{(1)} & \colhead{(2)} & \colhead{(3)} & \colhead{(4)} &
\colhead{(5)} & \colhead{(6)} \\
\colhead{Galaxy} & \colhead{Nucleus} & \colhead{$A_V$} &
\colhead{Nucleus (corr.)} & \colhead{X-ray} & \colhead{IRAS} \\
    & \colhead{[mJy]} & \colhead{[mags]} & \colhead{[mJy]} &
\colhead{[$\rm 10^{-15}\>W\>m^{-2}$]} & \colhead{[Jy]} 
}
\startdata
NGC\,253  & $400\pm100^a$ & $24\pm50\%^b$ & $1500^{+1500}_{-740}$ &
	${5.4^{+3}_{-2}}^{~c}$ & $25\pm7$ \\
NGC\,6946 & $300\pm 100^d$  & $2\pm50\%^e$ & $340^{+110}_{-85}$ &
	${1.4^{+0.5}_{-0.4}}^{~f}$ & $10\pm2$ \\
NGC\,7469 & $620\pm 80^g$  & $4\pm30\%^h$  & $770^{+110}_{-100}$ &
	${40^{+20}_{-13}}^{~i}$ \\
NGC\,7552 & $25\pm 5^j$  & $2\pm30\%^j$  & $28^{+6}_{-5}$ &
	${0.2^{+0.2}_{-0.1}}^{~k}$ & $2.6\pm0.3$ \\
M\,82	  & $200\pm100^l$  & $10\pm30\%^m$ & $350^{+180}_{-120}$ &
	${33^{+20}_{-14}}^{~n}$ & $72\pm14$ \\
NGC\,3690 & $700\pm100^o$ & $3.9\pm30\%^p$ & $870^{+140}_{-120}$ &
	${1.5^{+0.4}_{-0.3}}^{~q}$ & $3.8\pm0.8$ \\
NGC\,3256 & ${900^{+0}_{-500}}^r$ & $6.2\pm30\%^s$ &
	$1270^{+130}_{-460}$ & ${0.7^{+0.1}_{-0.1}}^{~t}$ & $3.2\pm0.7$ \\
NGC\,4051 & $330\pm50^u$ & $< 0.1^w$ & $330\pm 50$ &
	${16^{+8}_{-5.5}}^{~v}$ \\
NGC\,3227 & $225\pm50^u$ & $< 0.1^w$ & $225\pm 50$ &
	${25^{+13}_{-8}}^{~v}$ \\
NGC\,4151 & $1390\pm200^u$ & $20\pm40\%^x$ & $4200^{+2400}_{-1500}$ &
         ${230^{+120}_{-80}}^{~y}$ \\
\enddata

\tablecomments{(1): Galaxy name (2): N-band flux of the unresolved
nuclear component (3): estimated extinction towards the
$10\mum$ emitting region (see discussion in text) (4): foreground
extinction-corrected N-band flux of the unresolved nuclear source 
(5): absorption-corrected $(2-10)\kev$ X-ray flux (6): $12\mum$
IRAS flux within a $1.5\am$ diameter aperture.}

\tablerefs{
(a) \cite{boe98}; (b) \cite{sam94}; (c) \cite{pta97};
(d) \cite{tel93}, \cite{rie76}; (e) \cite{sch94}; (f) \cite{pta99a}, 
\cite{sch94};
(g) \cite{car88}, \cite{mil94}; (h) \cite{leb79}, \cite{ait81}; (i)
\cite{lei96};
(j) \cite{sch97}; (k) \cite{cha82}, \cite{mac81};
(l) \cite{tel91b}; (m) \cite{sat95}; (n) \cite{pta99}, \cite{mat99};
(o) \cite{ket97}; (p) \cite{alo00}; (q) \cite{zez98}, \cite{ris00};
(r) \cite{gra84}, \cite{rig99}; (s) \cite{doy94}; (t) \cite{mor99};
(u) \cite{dev87}, \cite{leb79}; (v) \cite{rey97}, \cite{geo98a}; (w) 
\cite{ho97}
(x) \cite{geo98b}; (y) \cite{yaq91}, \cite{geo98b}, \cite{wea94};
  	}
\end{deluxetable}

%%%%%%%%%%%%%%%%%%%%%%%%%%%%%%%%%%%%%%%%%%%%%%%%%%%%%%%%%%%%%%%%%%%%%%%%

\end{document}